\documentclass[apj]{emulateapj}

\usepackage{graphicx}
\usepackage{apjfonts}
\usepackage{mathptmx}



\def\***#1{{\sc #1}}
\def\plan#1{\relax}
\def\Plan#1{\relax}
\def\PLAN#1{\relax}

\def\lta{\mathrel{\spose{\lower 3pt\hbox{$\mathchar"218$}}
     \raise 2.0pt\hbox{$\mathchar"13C$}}}
\def\gta{\mathrel{\spose{\lower 3pt\hbox{$\mathchar"218$}}
     \raise 2.0pt\hbox{$\mathchar"13E$}}}

\shorttitle{}
\shortauthors{}

\def\mathnew{\mathsurround=0pt}

\def\simov#1#2{\lower .5pt\vbox{\baselineskip0pt \lineskip-.5pt
\ialign{$\mathnew#1\hfil##\hfil$\crcr#2\crcr\sim\crcr}}}

\begin{document}
\title{Statistical Evidence for Three classes of Gamma-ray Bursts}

\author{Tanuka Chattopadhyay$^1$, Ranjeev Misra $^2$, Asis Kumar Chattopadhyay$^3$ and Malay Naskar $^4$\\ }

\altaffiltext{1}{Shibpur Dinobundhoo College, 412/1 G.T. Road
(South), Howrah 711102, India, Email: tanuka@iucaa.ernet.in}

\altaffiltext{2}{Inter-University Centre for Astronomy and
Astrophysics, Post Bag 4, Ganeshkhind, Pune 411007. Email:
rmisra@iucaa.ernet.in}

\altaffiltext{3}{Department of Statistics,Calcutta University,35
Ballygunge Circular Road, Kolkata 700019, India.
Email:akcstat@caluniv.ac.in}

\altaffiltext{4}{NIRJAIT, Indian Council of Agricultural
Research,12, Regent Park, Kolkata 700 040, India. Email:
malaynaskar@yahoo.com}

\begin{abstract}
 Two different multivariate clustering techniques, the K-means partitioning
method and the Dirichlet process of mixture modeling, have been applied to
the BATSE Gamma-ray burst (GRB) catalog, to obtain the optimum number of
coherent groups. In the standard paradigm, GRB are classified
in only two groups, the long and short bursts. However, for both the
clustering techniques, the optimal number of classes was found to be three,
a result which is consistent with previous statistical analysis.
 In this classification,
the long bursts are further divided into two groups which are primarily
differentiated by their total fluence and duration and hence are named low
and high fluence GRB. Analysis of GRB with known red-shifts and spectral
parameters suggests that low fluence GRB have nearly constant isotropic
energy output of $10^{52}$ ergs while for the high fluence ones, the
energy output ranges from $10^{52}$ to $10^{54}$ ergs. It is speculated that the
three kinds of GRBs reflect three different origins: mergers of
neutron star systems, mergers between white dwarfs and neutron stars,
and collapse of massive stars

\end{abstract}


\section{Introduction}

Although it has now been well established that Gamma-Ray Bursts
(GRB) are of cosmological origin, their nature and source still
remains a mystery. Detailed observations and studies of their
afterglow emission have revealed important information regarding
the dynamic features and environments of these explosive events
(see \cite{Pir05} for a review). The detection of supernova light
curve in the afterglows of long duration nearby GRB has indicated
that a  fraction of the GRB occur during the the collapse of a
massive star (see \cite{Woo06} for a review). Other mechanism that could produce GRB are the
merger of compact objects like a pair of neutron stars or a
neutron star with a black hole \citep[e.g.][]{Pir92,Geh05,Blo06}. 
Thus GRB may be a heterogeneous
group and a  proper classification of the phenomena is crucial to
isolate and identify the possible different sources. Such a
classification will also enable the identification of spectral or
temporal correlations which may exist only for a particular class
of GRB.

In general, GRB have been classified into two groups of long ($> 2$ sec)
and short ($< 2$ sec) duration bursts.
This is based on visual inspection of the distribution of burst duration
 which clearly shows two peaks. Theoretically, this may be understood by identifying
long bursts with collapsing stars where the duration of the event is linked to
the dynamical collapse time-scale. On the other hand, the merger of two neutron
stars should occur on short timescales and hence may correspond to the
short duration bursts. In this scenario, the long duration bursts should
always be associated with a supernova explosion and occur in star burst regions, while
short bursts should have no relation to star burst regions and should have no associated
supernova. However, theoretically another mechanism to create a GRB could be the
merger of neutron stars with white dwarfs. These would be long duration bursts,
with no associated supernova and could be a significant fraction of the total
observed bursts \citep{Kin07}. Observationally, there have been some evidence that
there may be more than two classes of GRB. Some GRB have been recorded with low intrinsic
luminosity and do not comply with standard spectral relationships \citep{Saz04,Sod04}.
More compelling evidence was the absence of a
super-nova light curve in two nearby bursts, GRB 060614 and 060505, which suggested
that not all long duration bursts are due to  massive stellar collapses \citep{Geh06,Fyn06}
,although for GRB 060614 this result has been disputed by \cite{Sch06},
who claim that this GRB is not a nearby one.
 While confirmation of these results are awaited, they do highlight the need to examine
the possibility that there are more than two types of GRB.

The Burst and Transient Source Experiment (BATSE) on board the
COMPTON Gamma-Ray Observatory (CGRO) has provided spectral and
temporal information for more than 1500 GRB. Although BATSE
provides several spectral parameters, the bimodal distribution of
GRB is based on  univariate analysis i.e. only duration is
considered as a parameter \citep[e.g.][]{Dez92,Kou93}. There is a
claim that even such a univariate analysis supports the existence
of three classes \citep{Hor98}. Classification analysis taking
into account more observed parameters, i.e. a multivariate
analysis, was first undertaken by \cite{Fei98}. Subsequently,
different type of analysis were undertaken to classify GRB.
\cite{Bau94} used a neural network technique while \cite{Bag98}
have undertaken a factor analysis. Nonparametric hierarchical
clustering techniques have been used by \cite{Muk98} for 797 GRB
with six variables and by \cite{Bal01} for 1599 GRB with nine
variables. The results were confirmed for the complete BATSE GRB
catalog \citep{Hor02}. \cite{Hak03} used a unsupervised pattern
recognition algorithm while \cite{Hor06} have classified GRB by
fitting bivariate distributions to the observed duration and
hardness ratios. In all these cases, the authors have claimed the
existence of at least three classes of GRB although it is not
clear whether the different classifications found are consistent
with each other, primarily because of the different techniques and
the choice of different observed parameters and data sets.

These analyses are based on the observed properties of the GRB and
are hence subject to observational biases. In fact, \cite{Hak00}
argue that such classification techniques are significantly
hampered and the three classes found is probably due to  such
biases (see \cite{Hor06} for a counter argument). Indeed, a proper
classification should be based on intrinsic rather than observed
properties. However since the number of GRB with known red-shift
and uniformly measured temporal and spectral parameters is small,
such an exercise is presently not possible. Although it can be
argued that certain correlations between observed parameters (e.g.
the positive correlation between duration and flux) cannot be
entirely due to observational bias, it is difficult to ascertain
whether the quantitative relationships are not affected. A prudent
approach may be to treat such classification as indicative of the
nature of systems which should be corroborated by theoretical
expectations and further observations. Since theoretical
expectations are on intrinsic properties, there is a need to
estimate how these intrinsic properties cluster, given the results
of a classification based on observed ones. The classification
itself should be tested for robustness using different types of
schemes. The result of such an analysis may guide or be supported
by theoretical models. Moreover, the results can lay down the
broad framework and requirements of future observations which can
confirm or rule out the proposed  theoretical scenario.

In this work, we use two different multivariate clustering
techniques, the K-means partitioning method and the Dirichlet
process of mixture modeling, to classify GRB based on their
observed properties. These two schemes, which have not been used
before for GRB, have the advantage that they do not follow any
prior assumption about the number of homogeneous classes. The
optimum classification comes out of the process itself. The two
schemes allow for post classification discriminant analysis which
can be used to verify the acceptability of the classification by
computing classification/misclassification probabilities. More
importantly, a  GRB which is not in the original sample and which
has only a subset of the observed properties used for the
classification can be assigned a probability for it to be a member
of a certain class. Thus although our analysis has been based on
the BATSE catalog, GRB with known red-shifts (which have been
observed by other instruments) can be assigned such probabilities
and suitably classified. As we shall see, this not only provides
qualitative estimation of any possible  observational bias, but
can constrain the intrinsic properties of the different clusters.
We have selected 21 GRB with known redshifts and well constrained
spectral parameters for such classifications and obtained
constrains on the cluster's average luminosity.

In the next section, we briefly describe the two classification
schemes and present the result of the analysis on the BATSE
catalog. In \S 3, the classification obtained from the BATSE data
are used to classify GRB with known red-shifts and inferences are
made on the intrinsic properties of the different GRB groups. In
\S 4 the work is summarized and the main results are discussed.

\section{Clustering Analysis for GRB data}

The BATSE catalog provides temporal and spectral information for more than 1500
GRB. The parameters include, two measures of burst
durations, the times within which 50\% ($T_{50}$) and 90\%
($T_{90}$) of the flux arrive, three peak fluxes, $P_{64},
P_{256}, P_{1024}$ measured in 64, 256 and 1024 ms bins
respectively, four time integrated fluences $F_1-F_4$, in the
$20-50$, $50-100$, $100-300$ and $ > 300$ KeV spectral channels.
Many of the parameters are highly correlated and following previous
works \citep[e.g.][]{Muk98,Hak00} we use the following six parameter
set: log$T_{50}$,
 log$T_{90}$,  log$P_{256}$,  log$F_T$,  log$H_{32}$,  log$H_{321}$,
where  $F_T = F_1 + F_2 +F_3 +F_4$ is the total fluence while  $H_{32}=F_3/F_2$ and
$H_{321}=F_3/(F_1+F_2)$ are measures of spectral hardness. The sample consists of
$1594$ GRB that have non-zero detections of these parameters. We have not introduced
any completeness criteria (like a lower flux cutoff), since incompleteness primarily affects
the short duration bursts and hence is not expected to change the qualitative results
obtained.  We retain the $F_4$ flux (in the definition of $F_T$),
despite the uncertainties in its calibration and
sensitivity, because as we discuss later in \S 4, the $\gamma$-ray flux $> 300$ keV
is expected to have important spectral information. However, this fluence is not used
in the computation of spectral hardness.

\subsection{Partitioning (K-means clustering) method}

 Over the last several decades, different algorithms
have been developed for Cluster Analysis which is used to find groups in
a multivariate data set. The choice of a clustering algorithm
depends both on the type of data available and on the particular
purpose. Generally, clustering algorithms can be divided into two
principal types viz. partitioning and hierarchical methods.
A partitioning method constructs K clusters i.e. it classifies
 the data into K groups which together satisfy the requirement of
 a partition such that each group must contain at least one object and
 each object must belong to exactly one group. So there are at
 most as many groups as there are objects ($K <=n$). Two different
 clusters cannot have any object in common and the K groups
 together add up to the full data set.  The aim
 is usually to uncover a structure that is already present in the
 data.  On the other hand,
Hierarchical algorithms do not construct single partition with K
  clusters but they deal with all values of K in the same run. The
  extreme partitions with $K=1$ (all objects are
  together in one cluster)and $K = n$ (where each object forms a separate cluster)
is a part of the output. In between all values of
  $K=2,3,...n-1$ are covered in a kind of gradual transition. The
  only difference between $K=r$ \& $K=r+1$ is that one of the r
  clusters splits in order to obtain $r+1$ clusters or two of the
  $(r+1)$ clusters combined to yield $r$ clusters. In this method
  either one starts with $K=n$ and move hierarchically downwards
  where at each step two clusters are merged depending on
  similarity until only one is left i.e. $K=1$ (agglomerative) or
  the reverse way where one starts with $K=1$ and moves upwards
  where at each step one cluster is divided into two (depending on
  dissimilarity) until $K=n$ (divisive). Most of the previous
  works  on GRB \citep[e.g.][]{Muk98,Hak00,Bal01} have been based on
hierarchical clustering. However, for GRB classification, a
 partitioning  method may be more applicable because
  (a) it tries to select the best clustering with K
  groups which is not the goal of a hierarchical method,
  (b) a hierarchical method can never repair what was done in
  previous steps and (c) partitioning methods are designed to group items rather than
  variables into a collection of K clusters.

In this work, we apply the K- means method of \cite{Mac67} which is probably the
 most widely technique, to the BATSE catalog.  For this method,
the optimum value of K  can be obtained in different ways \citep{Har75}.
This is done by computing for each cluster formation ( i.e for number
of clusters $K = 2,3,4..$) a distance measure
 $d_{K} = (1/p) min_{x} E[( x_{K} - c_{K} )^{'}(x_{K} - c_{K})]$ which is defined
as the distance of
the $x_{K}$ vector (values of the parameters) from the center of a cluster
$c_{K}$ (which is estimated as mean value). $p$ is the order of the
$x_{K}$ vector, i.e. the number of parameters which for our case is six.
If $d_{K}^{\prime}$ is the estimate of $d_{K}$ at the
$K^{th}$ point, then $d_{K}^{\prime}$ is the minimum achievable
distortion associated with fitting K centers to the data.
A natural way of choosing the number of clusters is to plot
$d_{K}^{\prime}$ versus K and look for the resulting distortion
curve. This curve will monotonically decrease with increasing K, till
K is greater than true number of clusters, after which the curve will
level off with a smaller slope. This is expected since adding more clusters
beyond the true number, will simply create partitions within a group.
According to \cite{Sug03} it is more illustrative to consider the
transformation of the distortion curve to an appropriate negative power,
$J_{K} = (d_{K}^{\prime-(p/2)}  - d_{K-1}^{\prime-(p/2)}$), which will exhibit a
sharp "jump" when K equals the true number of clusters.
The optimum number of clusters is the value of  K at which the
distortion curve levels off as well as its value associated with
the largest jump for the transformed curve.

\begin{figure}
\begin{center}
{\includegraphics[width=1.0\linewidth,angle=0]{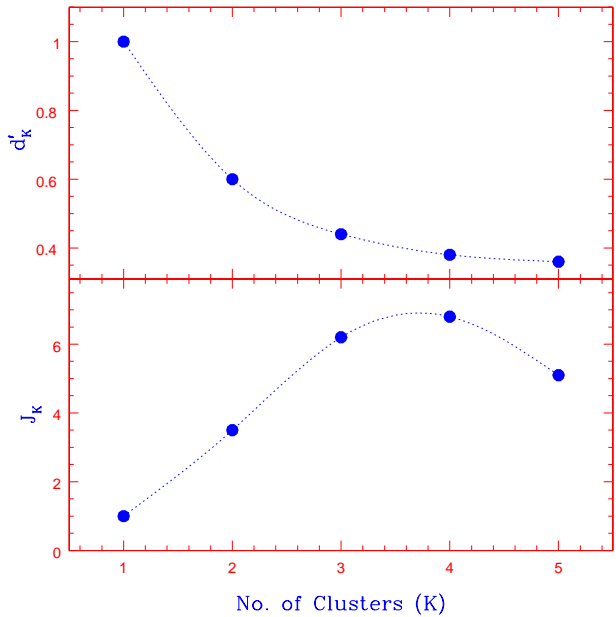}}
\end{center}
\caption{The distortion curve $d_{K}^{\prime}$ and the transformed jump curve $J_K$ for
different number of clusters. The largest value of $J_K$ and the leveling of $d_{K}^{\prime}$
indicates that for the K-means clustering technique the optimal number of clusters is three. }
\label{f1}
\end{figure}

Fig \ref{f1} shows the distortion curve and transformed jump curve
for the analysis of the BATSE data. The leveling of the distortion
curve and the shape of the jump function strongly suggests that
the optimum number of clusters is greater than two and is likely
to be three. The group means and the standard errors for the six
parameters for three cluster classification, are tabulated in
Table \ref{t1}. Cluster I with 423 members has an average
$<T_{90}> \sim 0.5$ s can be clearly identified with the short
duration bursts. The long duration bursts are cleanly separated
into two clusters (clusters 2 and 3) with 622 and 549 members.

\begin{deluxetable} {lrrr}
\tablewidth{0pt}
\tablecaption{Average Cluster properties based on K-means classification}
\tablehead{
\colhead{Parameters} & \colhead{Cluster I} & \colhead{Cluster II} &
\colhead{Cluster III} }
\startdata
$T_{50}$ (sec) &$0.19\pm0.01$& 5.37$\pm$ 0.25&22.9$\pm$ 1.0\\
$T_{90}$ (sec)&0.50$\pm$ 0.02&15.85$\pm$ 0.73&63.1$\pm$ 2.9\\
$P_{256}$ (\#/cm$^2$/sec) &1.66$\pm$ 0.08& 1.26$\pm$ 0.06& 2.88$\pm$ 0.13\\
$F_T$ ($\times 10^{-6}$ ergs/sec)&$0.62\pm{0.04}$& $2.34\pm{0.11}$&$17.8\pm{0.8}$\\
$H_{32}$&5.50$\pm$ 0.13& 2.45$\pm$ 0.06& 3.16$\pm$ 0.07\\
$H_{321}$&3.39$\pm$ 0.08& 1.32$\pm$ 0.03& 1.78$\pm$ 0.04\\
\enddata
\tablecomments{Errors quoted are standard errors. The number of members are
423, 622 and 549 for Clusters I,II and III respectively.}
\label{t1}
\end{deluxetable}

Once the optimum classification (clustering) is obtained, using a
process called Discrimination Analysis \cite{Joh96}, one can
verify the acceptability of the classification by computing
classification/misclassification probabilities for the different
GRB.  Although the K-means clustering method is purely a data
analytic method, for classification it may be necessary to assume
that the underlying distribution is Multivariate Normal. In this
standard procedure, using the probability density functions in
parameter space for the different clusters, one can assign an
object (in this case a GRB) to be a member of a certain class. If
the original classification was robust, then every GRB
should be classified again as a member of the same class that it was
before. If a
significant number of objects are not reclassified then that would
mean that the original classification was not stable and hence not
trustworthy. Table \ref{t2} show the result of a Discrimination
Analysis, where the columns represent how the GRB of a cluster
were assigned by the analysis. The
 fraction of correct classification is $0.954$ which implies that the classification is
indeed robust.

\begin{deluxetable} {lccc}
\tablewidth{0pt}
\tablecaption{Discriminant Analysis for the  K-means classification}
\tablehead{
\colhead{} & \colhead{Cluster I} & \colhead{Cluster II} &
\colhead{Cluster III} }
\startdata
Cluster $I^*$ &417& 28&  0\\
Cluster $II^*$ &  6&578& 23\\
Cluster $III^*$ &  0& 21&526\\
&&&\\
Total         &423&622&549\\
\enddata
\tablecomments{Clusters I,II and III, are the clusters obtained from the K-means classification.
Clusters $I^*,II^*$ and $III^*$ are the clusters to which the GRB were assigned by the Discriminant
analysis.}
\label{t2}
\end{deluxetable}

\subsection{Dirichlet process model based clustering}

The standard approach to  model-based clustering analysis, is
based on modeling by finite mixture of parametric distributions.
For example, \cite{Muk98} used such a model based approach to analyze
GRB data where they assumed that the GRB population
consists of mixture of multivariate Gaussian classes. The number
of classes is, however, determined from an initial classification
method (e.g. via agglomerative hierarchical clustering).  The
Dirichlet process model based clustering is more general and avoids the
assumption of known number of possible classes. Since this method
is less commonly used as compared to the K-means technique, we describe here
the basic concept on the analysis in more detail.

The Dirichlet process avoids a prior assumption of the number of classes
by applying a Bayesian
nonparametric modeling of the unknown distribution for the multi component data. In this
particular case the six component GRB data can be represented by
$x_i=($log$T_{50},$log$T_{90},$log$F_{T},$log$P_{256},$log$H_{321},$log$H_{32})^\prime,i=1,2,\ldots,n$.
More specifically, $x_i$ is assumed to follow a multivariate normal
distribution whose mean vector is generated from a Dirichlet
Process (DP). Following \cite{Esc95}, the method is
best conceptualized by representing the model as
\begin{eqnarray}\label{eq:model}
x_i|\mu_i, \Sigma & \sim & \hbox {MVN}(\mu_i,\Sigma), \;\;\;\; i=1,2,\ldots,n; \nonumber \\
\mu_i|G & \sim & G ; \;\;\;\;\;\; G \sim DP(\alpha G_0)
\end{eqnarray}
where MVN means multivariate normal distribution, $G$ is a
discrete measure of the unknown  distribution, $\alpha$ is the precision
parameter and $G_0$ is a known base measure distribution. Since
$G$ is discrete, there can be ties among the $\mu_i$'s , which can
also be seen from polya urn representation of \cite{Bla73} as
\begin{equation}\label{eq:bm}
\mu_i|\mu_1,\mu_2,\ldots,\mu_{i-1} \sim
\frac{\alpha}{\alpha+i-1}G_0 + \frac{1}{\alpha+i-1}
\sum_{h=1}^{i-1} \delta(\mu_h)
\end{equation}
 where $\delta(x)$ is the distribution concentrated at the single
 point $x$. It is evident from Eqn. (\ref{eq:bm}) that $\mu_i$ are
marginally sampled from $G_0$ with positive probability and that some
of the $\mu_i$'s are identical. Thus, a partition of
$S=\{1,2,\ldots,n\}$ can be formed by defining classes under the
relation that $\mu_i$ belongs to the $j^{th}$ class if and only if
$\mu_i=\mu_j , j=1,2,\ldots,k$, $k$ being the number of distinct
$\mu_i$'s $, i=1,2,\ldots,n$. This induces a certain posterior
distribution of $S$ and a posterior inference  can then be used to
provide clustering procedure. There are various algorithms
available to obtain the posterior partitions of $S$ which are
useful for making inferences on clustering of $x_1,x_2, \ldots,
x_n$. We implement  the independent and identically distributed
Weighted Chinese Restaurant (iidWCR) algorithm \citep[see][]{Ish02}
which comes from its use of the partition
distribution of $S$. Let $p=\{C_1,C_2,\ldots,C_{n(p)}\}$ be a
partition of size $n(p)$ of $S$, where each $C_j$ contains $e_j$
elements. Assuming $G_0 $ as the multivariate normal with mean
vector $m$ and covariance matrix $B_0$ and denoting
 $N_p(x;\mu,\Sigma)=(2\pi)^{-\frac{p}{2}}|\Sigma|^{-\frac{1}{2}}
 exp[-\frac{1}{2}(x-\mu)^\prime \Sigma^{-1}(x-\mu)]$
as the density of a $p$-component multivariate normal
distribution, the iidWCR algorithm for inducing posterior
partition of $S$ consists of following steps:
\begin{description}
\item[\emph{Step 1:}] Assign $p_1 = \{1\}$ and the corresponding
importance weight $\lambda (1)=N_6(x_1;m,\Sigma_0+B_0)$ where
$\Sigma_0$ is the initial estimate of $\Sigma$.

\item[\emph{Step r:}] Given $p_{r-1}$, compute $\Sigma_{r-1}$ from
$x_1,x_2,\ldots, x_{r-1}$. Create $p_r$ by assigning label $r$ to
a new set with probability $\frac{\alpha}{(\alpha+r-1)\lambda(r)}
\times N_6(x_r;m,\Sigma_{r-1}+B_0)$. Otherwise, assign label $r$
to an existing set $C_{j,r-1}$ with probability $
\frac{e_{j,r-1}}{(\alpha+r-1)\lambda(r)} \times
N_6(x_r;\mu_{j,r-1}, \Sigma_{j,r-1})$ where
$\Sigma_{j,r-1}=(B_0^{-1} +e_{j,r-1}\Sigma_{r-1}^{-1})^{-1}$ and
$\mu_{j,r-1}=\Sigma_{j,r-1}(B_0^{-1}+e_{j,r-1}\Sigma_{r-1}^{-1}\overline{x}_{j,r-1})$.
Note that $e_{j,r-1}$ and $\overline{x}_{j,r-1}$ are the  number
of elements and observed mean in $C_{j,r-1}$ respectively and
$\lambda(r)$ is the normalizing constant.
\end{description}
Running step 1 followed by step $r$ for $r=2,3,\ldots,n$ gives a
draw from posterior partition of $S$. This $n-step$ draw, in fact,
provides an iid sample from WCR density given by
\begin{equation}
g(p) = \frac{f(x|p)\pi(p)}{\Delta(p)}\nonumber
\end{equation}
where $\pi(p)$ is the prior density of $p$, f(.) is the density of
$x$ and $\Delta(p)=\lambda(1) \times \lambda(2) \times \ldots
\times \lambda(n)$ is the importance weight. Repeating the above
algorithm $B$ times, one can obtain $p^1,p^2,\ldots,p^B$ iid
sample observations from posterior partition of $S$. Based on
these sample observations, Monte Carlo method can be devised to
estimate $E\{n(p)\}$, the expected number of clusters, as
\begin{equation}\label{eq:est-cluster}
\widehat{E\{n(p)\}} \approx \frac{\sum_{b=1}^B n(p^b)
\Delta(p^b)}{\sum_{b=1}^B  \Delta(p^b)}
\end{equation}
The key advantages of using this Dirichlet process model-based
clustering are that  the underlying distribution of $x_i$'s and
the number of clusters are unknown. Moreover, one can provide an
estimate of the expected number of clusters by using Eqn.
(\ref{eq:est-cluster}).

\begin{deluxetable} {lrrr}
\tablewidth{0pt}
\tablecaption{Average Cluster properties based on the Dirichlet Mixture Modeling method }
\tablehead{
\colhead{Parameters} & \colhead{Cluster I} & \colhead{Cluster II} &
\colhead{Cluster III} }
\startdata
$T_{50}$ (sec) &0.31$\pm$ 0.02& 6.76$\pm$ 0.31&16.22$\pm$ 1.50\\
$T_{90}$ (sec)& 0.45$\pm$ 0.03&19.05$\pm$ 0.88&43.65$\pm$ 3.02\\
$P_{256}$ (\#/cm$^2$/sec) &1.66$\pm$ 0.08& 1.35$\pm$ 0.03& 4.79$\pm$ 0.33\\
$F_T$ ($\times 10^{-6}$ ergs/sec)&$0.89\pm{0.06}$& $3.46\pm{0.08}$&$18.2\pm{0.2}$\\
$H_{32}$&4.68$\pm$ 0.22& 2.82$\pm$ 0.06& 3.31$\pm$ 0.15\\
$H_{321}$&2.75$\pm$ 0.13& 1.58$\pm$ 0.04& 1.86$\pm$ 0.09\\
\enddata
\tablecomments{Errors quoted are standard errors. The number of members are
409, 892 and 293 for Clusters I,II and III respectively.}
\label{t3}
\end{deluxetable}

For fitting the model, we used $\alpha=1.0$ and a flat
prior $G_0 \sim N_6(0,\sigma^2\mathbf{I})$ with $\sigma^2=1000$.
The initial value of $\Sigma$, $\Sigma_0$ is obtained as the
sample covariance matrix. We applied the iidWCR algorithm for
$B=1000$ to obtain the estimate of the number of clusters.
Three classes were obtained consistent with the results from
the K-means technique.    In Table \ref{t3} the mean values of
the parameters with errors are tabulated. The values are consistent
with those found from the K-means method.
The number of members of cluster II, 892 is somewhat larger
than what was found by the K-means method, 622, but considering the different nature and
approach of the two techniques, such differences are perhaps expected. In summary,
these  two independent
clustering techniques indicate that there are three classes of GRB with
qualitatively similar properties.

\section{Classification of GRB with known red-shift}

\begin{figure}
\begin{center}
{\includegraphics[width=1.0\linewidth,angle=0]{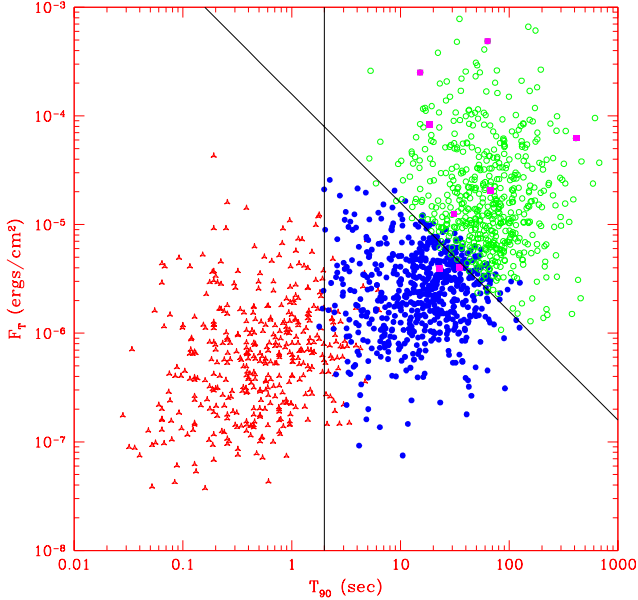}}
\end{center}
\caption{The fluence $F_T$ in ergs/cm$^2$ versus the duration $T_{90}$ in seconds for members
of Cluster I (triangles), II (solid circles) and III (open circles) from the 
K-means clustering method. The solid lines represent
$T_{90} = 2$s and $F_T = 1.6 \times 10^{-4}/T_{90}$ ergs/cm$^2$ which qualitatively 
separate the three groups. The solid squares represent eight GRB detected by 
BATSE for which redshifts are also measured \citep[e.g.][]{Bag03}. }
\label{f2}
\end{figure}

Although the classification described in the previous section is based on six GRB parameters
the segregation of the classes can be visualized using the total fluence, $F_T$ and the
duration, $T_{90}$. This is illustrated in Fig (\ref{f2}) which shows  $F_T$ versus
$T_{90}$ for the members of the three clusters obtained using the K-means technique. The
Dirichlet process model gives qualitatively similar results. The solid line representing
$T_{90} = 2$ sec, differentiates the members of Cluster I (marked
by triangles) with those of Cluster II (marked by filled circles). Thus Cluster I is consistent
with the standard classification of short duration bursts. The standard long duration bursts
(with $T_{90} > 2$ s) are further classified into two groups with one of them (members
of Cluster III, marked using open circles) having typically higher fluence. Thus we have named
members of Cluster II and III as low and high  fluence GRB. The
solid line representing $F_T = 1.6 \times 10^{-4}/T_{90}$ ergs/cm$^2$ qualitatively separates two groups. There are eight GRB detected by BATSE for which
there are redshift estimates \citep[e.g.][]{Bag03}. These are marked by
squares in Figure 2. Six of them are in Cluster III while two are close to
the demarking line. One of these GRB (980425) is at a low redshift 
($z = 0.0085$) and is associated with a supernova. 

\begin{figure}
\begin{center}
{\includegraphics[width=1.0\linewidth,angle=0]{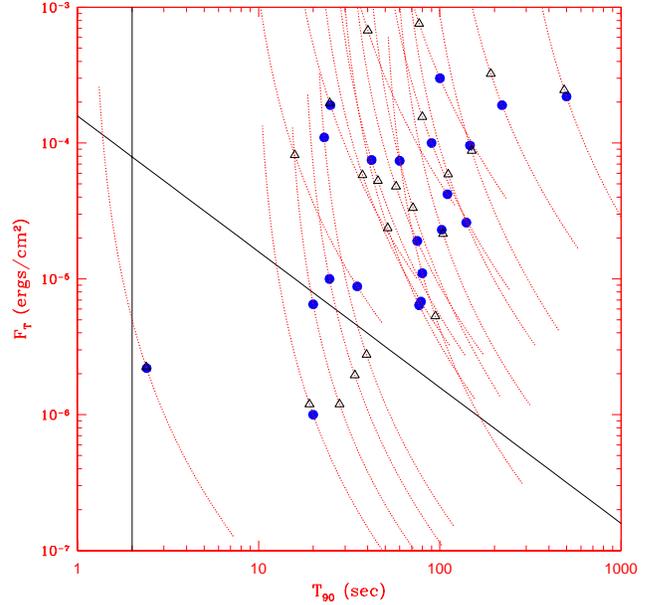}}
\end{center}
\caption{ The fluence $F_T$ in ergs/cm$^2$ versus the duration $T_{90}$ in
seconds for $21$ GRB (filled circles) with known redshifts taken from
\cite{Ghi04} and references therein.  The solid lines represent $T_{90} = 2$s and $F_T =
1.6 \times 10^{-4}/T_{90}$ which qualitatively distinguish the
three clusters (Fig \ref{f2}). For each GRB, the dotted lines
represent the predicted fluence and duration if it was located at
a redshift range $0.1 < z < 5$. The open triangles represent the
predicted fluence and duration if the GRB were located at
redshift $z = 1$. These predicted values allow for the
classification of the GRB into members of Cluster II and III. }
\label{f3}
\end{figure}

To identify a GRB with  a known red-shift as a member of a
cluster, broad band coverage of the prompt emission is required in
order to correctly estimate the total fluence $F_T$, which BATSE
would have observed for the burst. This is particularly important,
when the peak of the energy spectrum of a GRB is at high energies
$> 300$ keV. \cite{Ama02} analyzed GRB with known red-shifts and
well constraned spectral parameters over a broad energy range and
discovered that the intrinsic (i.e. red-shift corrected) peak of
the energy spectrum, $E_p$ correlates with the isotropic energy
output, $E_{iso}$. Apart from being a stringent condition and test
for any theoretical model that describes the GRB prompt emission,
this empirical relation highlights the possibility that GRB can be
used to probe and constrain the expansion of the Universe at early
times. \cite{Ghi04} added more GRB to the sample and  found that
the the beaming corrected luminosity, $E$ has a tighter
correlation with $E_p$ than the isotropic one. Nevertheless, there
is still significant dispersion in the relationship which needs to
be explained. In order to see how the classification  found in
this work affects such relationship, we  have used 21 GRB listed
in Table 1 of \cite{Ghi04} that have well measured temporal and
spectral parameters. The total fluence $F_T$ versus the duration
$T_{90}$ for these GRB are plotted in Fig. (\ref{f3}) (filled
circles). Overlaid on the plot are the two solid lines
 $F_T = 1.6 \times 10^{-4}/T_{90}$ ergs and $T_{90} = 2$s which qualitatively segregate the three
clusters (Fig \ref{f2}). Most of the GRB have high fluence and are of long duration (consistent with them
being members of Cluster III) which is probably a selection effect.  Two of the GRB have $F_T < 1.6 \times 10^{-4}/T_{90}$ ergs/cm$^2$
and are probably members of Cluster II. However, there are three GRB with
 $F_T \sim 1.6 \times 10^{-4}/T_{90}$ ergs/cm$^2$ and hence there is an ambiguity about their classification.
For each GRB, a corresponding dotted line is plotted in
Fig (\ref{f3}) which shows the variation of the observed $F_T$ versus $T_{90}$ if the same GRB was
located at different red-shifts. The lines are drawn for a red-shift range $0.1 < z < 5.0$ for a
$\Lambda$ CDM cosmology with $\Omega_\Lambda = 0.7$ and Hubble parameter $H = 65$ km/s/Mpc.
It is interesting to note that the red-shift trajectories for the high fluence GRB in general do
not cross the $F_T = 1.6 \times 10^{-4}/T_{90}$ line. In other words, if these GRB were located
at a wide range of red-shifts, their observed fluence would have been $F_T > 1.6 \times 10^{-4}/T_{90}$
and hence they would have been classified as members of Cluster III. On the other hand there are six
GRB whose red-shift trajectories mostly lie below the de-marking line. Their  observed fluence would be
$F_T < 1.6 \times 10^{-4}/T_{90}$ for a wide range of red-shifts and hence they would be classified
as members of Cluster II. The open triangles in the Figure mark the positions of the GRBs if they were
all located at a red-shift, $z = 1.0$. In this representation, the GRB are more clearly segregated into Clusters
II and III with five of them having a  predicted fluence less
than  $F_T > 1.6 \times 10^{-4}/T_{90}$ with the others being significantly brighter. This strongly suggests
that the classification described in this work is not due to observational bias arising from the use
of observed parameters instead of intrinsic ones.

\begin{figure}
\begin{center}
{\includegraphics[width=1.0\linewidth,angle=0]{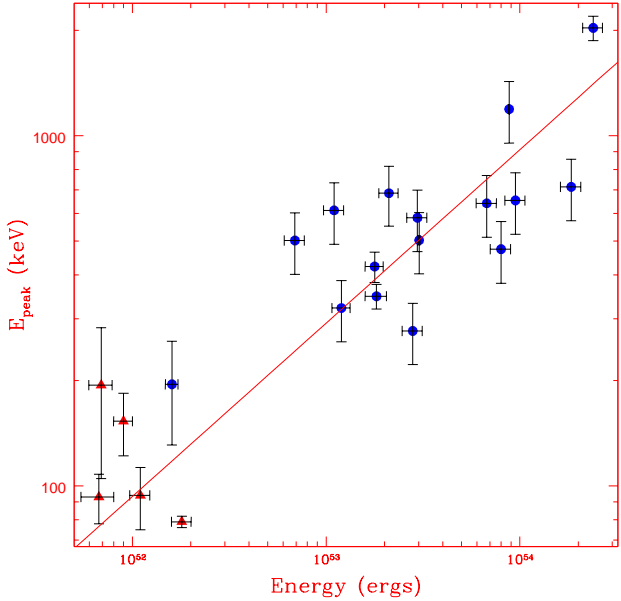}}
\end{center}
\caption{ The intrinsic (red-shift corrected) peak of the energy flux, $E_{peak}$ versus the
isotropic energy output for 21 GRB with well defined spectral parameters \citep{Ama02,Ghi04}.
GRB that are members of Cluster II (triangles) have isotropic energy output of nearly
$10^{52}$ ergs, while members of Cluster III (circles) have a much wider range of energy
output.}
\label{f4}
\end{figure}

We classify the 21 GRB according to their predicted observed fluence and duration if they were situated
at $z = 1$. In this scheme, five GRBs
are classified as members of Cluster II while the remaining 18 are identified as
members of Cluster III. Fig (\ref{f4}) shows the variation of the intrinsic energy peak $E_{peak}$ versus
the isotropic energy realized, $E_{iso}$, which is the correlation discovered by \cite{Ama02}. GRB identified
as Cluster I are marked as triangles while those belonging to Cluster III are represented by filled circles.
GRB belonging to Cluster II all
have an isotropic energy output close to $\sim 10^{52}$ ergs, while those belonging  to Cluster III span
a much larger range of energy $10^{52-54}$ ergs and roughly follow the $E_p-E_{iso}$ correlation. Although the
number of GRB in this sample is small, this segregation of the GRB in intrinsic parameter space is a
another indication that the classification is robust and perhaps not due to observational bias.
Most of the members of Cluster III have a rest frame peak energy $E_{peak} > 300$ keV and hence would have
had significant flux in the highest energy channel of BATSE, i.e. $F_4$. Thus, the retention of $F_4$ in
the classification analysis of \S 2 is important even though there is systematic  uncertainty in
the measured value of the fluence. Indeed, if $F_4$ is not taken into account the evidence for  three
clusters in the BATSE sample decreases.

\section{Summary and Discussion}

Two multivariate clustering techniques, the K-means partitioning
method and the Dirichlet process of mixture modeling, have been applied for the first time to
the BATSE Gamma-ray burst (GRB) catalog. These two schemes do not make any a priori assumptions
about the number of clusters, but instead provide quantitative estimate of the optimal number of
groups. The jump curve for the K-means partitioning method suggests that this optimal number is
three which is further supported with the value of  the expected number of clusters, $E\{n(p)\}$, obtained
using Dirichlet process of mixture modeling. The two techniques group the GRBs in qualitatively similar
classes, which can be described as short bursts ($T_{90} < 2$ s, Cluster I), long duration, low fluence bursts
($F_T < 1.6 \times 10^{-4}/T_{90}$ ergs/cm$^2$, Cluster II) and long duration, high fluence bursts ($F_T > 1.6 \times 10^{-4}/T_{90}$ ergs/cm$^2$, Cluster III).

To estimate how such a classification, based on observed spectral and temporal parameters,
 can arise from intrinsic GRB properties, a sample of 21 GRB with known red-shifts and well constrained
spectral parameters,  are classified within
this scheme. The observed total fluence, $F_T$ and duration $T_{90}$ for
these GRBs if they were located at different red-shifts ($0.1 < z < 5$) were estimated. For 16 of the
21 GRB, the estimated fluence would have satisfied $F_T > 1.6 \times 10^{-4}/T_{90}$ ergs/cm$^2$
for nearly the entire
range of red-shift space and hence they were classified as high fluence bursts.
This invariance in red-shift, indicates that the  classification scheme is not strongly effected
by observational bias and by the use of observed parameters instead of intrinsic ones. For
five GRB, classified as low fluence bursts, the predicted fluence   $F_T < 1.6 \times 10^{-4}/T_{90}$
for a significant fraction of the red-shift space, which again signifies the physical nature
of the classification. Based on the classification of these GRB with known red-shift, it can
be inferred that the low fluence GRB have a  nearly constant isotropic Energy output of $10^{52}$ ergs
and have an intrinsic (red-shift corrected) duration of $T_{90} \sim 2$-$30$ secs.
On the other hand, the high fluence GRB (Cluster III)
have a much wider range of isotropic energy output $10^{52-54}$ ergs and a corresponding wide
range of intrinsic durations $10$ - $500$ secs.

We note with caution that the number of GRB with known red-shifts, used for this analysis
is small and a much larger sample is required before concrete conclusions can be drawn. It is
also important, for this analysis,  to have well constrained spectral parameters of these GRB.
In particular, the peak of the energy fluxes, $E_p$ are required to be well estimated
and since for high energetic
sources $E_p > 300$ keV, it is imperative to have well calibrated high energy information. Indeed, if the
highest energy channel of the BATSE measurement is not taken into account,
the significance of the classification is smaller.

The classification presented here, needs to be supported by
theoretical considerations. It is tempting to identify the low
fluence GRB with neutron star-white dwarf mergers \citep{Kin07}
and the higher fluences ones with massive stellar collapse. The
near constancy of the isotropic energy output of low fluence
bursts, seem to be consistent with them being neutron star-white
dwarf mergers. Since both neutron stars and white dwarfs do not
have significant mass variations, their initial conditions for the
binary merger could be similar, leading to the nearly constant
energy output. Moreover, their merger time may also be typically
smaller than massive stellar collapse time-scales, which is
consistent with the shorter intrinsic duration $2-30$ s, found in
this work. On the other hand, the energy output and duration of
GRB induced by massive stellar collapse may depend on the mass and
size of the progenitor which is consistent with the variation
inferred for high fluence bursts. The present observational
evidence for such a model is not clear. Evidence for supernova
light curves have been detected in GRB with different energy
output, including some low luminosity ones, e.g. GRB 0311203, $E
\sim 3 \times 10^{49}$ ergs \citep{Mal04}, the nearby GRB 060614,
for which no supernova was detected also had a low isotropic
energy output of $10^{51}$ ergs. Such low energy output are not
represented in the 21 GRB with well constrained spectral
parameters used in this analysis which all have energies $>
10^{52}$ ergs. These rare GRB (since they have to be located
relatively nearby to be detected) may not represent a significant
fraction of the BATSE catalogue. BATSE did detect GRB 980425 which
is at low redshift ($z = 0.0086$) and is associated with a supernova. 
At this redshift, the GRB would be a borderline case between the high and 
low fluence GRB (Figure 2). Thus the interpretation of the
two different classes of long bursts as being due to stellar
collapse and white dwarf-neutron star mergers, is speculative and
more quantitative theoretical predictions and observational
evidences are required before a definite conclusion can be made.

\acknowledgements

TC and AKC thank the IUCAA associateship program for
support

\end{document}